\documentclass[10pt,a4paper]{article}
\usepackage{geometry}
\usepackage[utf8]{inputenc}
\usepackage{amsmath}
\usepackage{amsfonts}
\usepackage{amssymb}
\usepackage{graphicx}
\usepackage{algorithm}
\usepackage[noend]{algpseudocode}
\usepackage{mathrsfs}
\usepackage{amsmath}
\usepackage{amsfonts}
\usepackage{amssymb}
\usepackage{dsfont}
\usepackage{csquotes}
\usepackage{rotating}

\usepackage{multirow}
\usepackage{xcolor}
\usepackage{cancel}
\usepackage{tikz}
\usetikzlibrary{calc}
\usepackage[hidelinks]{hyperref}
\usepackage[symbols,nogroupskip,nonumberlist,sort=use]{glossaries-extra}
\glsdisablehyper

\def\ie{i.\,e.}
\newcommand\vek[1]{\ensuremath{\vec{#1}}}

\newcommand\Prob[1]{\ensuremath{P\left(#1\right)}}
\newcommand\condP[2]{\ensuremath{\Prob{#1 \vert #2}}}
\newcommand\Poissonian[2]{\ensuremath{e^{#2} \frac{#2^{#1}}{#1!}}}
\newcommand\Gauss[2]{\ensuremath{\mathcal{G}\left(#1,#2\right)}}
\newcommand\Gaussian[2]{\ensuremath{
    \frac{1}{\sqrt{2\pi \left\vert #2 \right\vert}}
    \exp\left(-\frac{1}{2}{#1}^\dagger {#2}^{-1} #1\right)
}}
\newcommand\ExpV[1]{\ensuremath{\left\langle #1 \right\rangle}}
\newcommand\PriorCov[1]{\ensuremath{S_{#1}}}

\def\timevar{\ensuremath{\tau}}
\def\timeunit{\ensuremath{\timevar_\Delta}}

\def\pprime{\ensuremath{{\prime \prime}}}

\glsxtrnewsymbol[description={log concentration}]{tildec}{\ensuremath{\mu}}

\glsxtrnewsymbol[description={coldness (thermodynamic beta)}]{ctb}{\ensuremath{\beta}}

\glsxtrnewsymbol[description={volume}]{vol}{\ensuremath{V}}
\def\vol{{\ensuremath{\gls{vol}}}}

\glsxtrnewsymbol[description={reactor volume}]{volR}{\ensuremath{\vol_\text{R}}}

\glsxtrnewsymbol[description={Avogadro constant}]{avo}{\ensuremath{N_\text{A}}}

\glsxtrnewsymbol[description={Gibbs free energy}]{gibbs}{\ensuremath{\mathcal{G}}}
\def\gibbs{{\ensuremath{\gls{gibbs}}}}

\glsxtrnewsymbol[description={concentration}]{c}{\ensuremath{c}}
\def\c{{\ensuremath{\gls{c}}}}

\glsxtrnewsymbol[description={motif concentration}]{mc}{\ensuremath{c}}
\def\mc{{\ensuremath{\gls{mc}}}}

\glsxtrnewsymbol[description={complex concentration}]{cc}{\ensuremath{\mathfrak c}}
\def\cc{{\ensuremath{\gls{cc}}}}

\glsxtrnewsymbol[description={motif concentration vector}]{mcv}{\ensuremath{\vek \mc}}
\def\mcv{{\ensuremath{\gls{mcv}}}}
\glsxtrnewsymbol[description={motif concentration trajectory}]{mct}{\ensuremath{\mcv\left(\timevar\right)}}
\def\mct{{\ensuremath{\gls{mct}}}}

\glsxtrnewsymbol[description={complex concentration vector}]{ccv}{\ensuremath{\vec \cc}}

\glsxtrnewsymbol[description={concentration of a single particle}]{concentrationOfASingleParticle}{\ensuremath{\cc_\Delta}}

\glsxtrnewsymbol[description={creation time of first dimer}]{ctofd}{\ensuremath{\timevar_\text{dim}}}

\glsxtrnewsymbol[description={Boltzmann constant}]{kB}{\ensuremath{k_\text{B}}}

\glsxtrnewsymbol[description={Temperature}]{T}{\ensuremath{T}}

\def\nuclvar{\ensuremath{x}}

\newcommand\komplement[1]{\ensuremath{{#1}^*}}
\glsxtrnewsymbol[description={complement of nucleotide $\nuclvar$}]{komplementx}{\ensuremath{\komplement{\nuclvar}}}

\def\motifvar{\ensuremath{m}}
\newcommand\mkomplement[1]{\ensuremath{{#1}^\dagger}}
\glsxtrnewsymbol[description={complement of a motif $\motifvar$}]{mkomplementm}{\ensuremath{\mkomplement{\motifvar}}}

\glsxtrnewsymbol[description={strand number}]{strandnumber}{\ensuremath{n}}

\glsxtrnewsymbol[description={segment length}]{segmentlength}{\ensuremath{l}}

\glsxtrnewsymbol[description={complex length}]{complexlength}{\ensuremath{L}}

\glsxtrnewsymbol[description={free length}]{freelength}{\ensuremath{\widetilde L}}
\def\freelength{{\ensuremath{\gls{freelength}}}}
\glsxtrnewsymbol[description={free length for rotation invariant complex}]{symfreelength}{\ensuremath{\freelength^\prime}}

\glsxtrnewsymbol[description={remaining length distributions}]{rld}{\ensuremath{\sigma^\prime}}

\glsxtrnewsymbol[description={remaining length distributions correction}]{rldc}{\ensuremath{\sigma^\pprime}}

\glsxtrnewsymbol[description={correct remaining length distributions}]{crld}{\ensuremath{\sigma}}

\glsxtrnewsymbol[description={preliminary complex number for given length and strand number}]{pcnln}{\ensuremath{\nu^\prime}}

\glsxtrnewsymbol[description={(correct) complex number for given length and strand number}]{cnln}{\ensuremath{\nu}}

\glsxtrnewsymbol[description={reference concentration in $\text{mol}/\text{L}$}]{c0}{\ensuremath{\cc_0}}

\glsxtrnewsymbol[description={alphabet}]{alphabet}{\ensuremath{\mathcal{A}}}
\def\alphabet{{\ensuremath{\gls{alphabet}}}}
\glsxtrnewsymbol[description={extended alphabet}]{ealphabet}{\ensuremath{\alphabet_0}}

\glsxtrnewsymbol[description={number of letters in the alphabet}]{nalphabet}{\ensuremath{|\alphabet |}}

\glsxtrnewsymbol[description={alphabet order}]{alphabetord}{\ensuremath{\mathcal{N}_\mathcal{A}}}

\def\letterI{\ensuremath{\mathrm{X}}}
\def\letterII{\ensuremath{\mathrm{Y}}}

\glsxtrnewsymbol[description={total mass, \ie number of nucleotides}]{tm}{\ensuremath{M}}

\glsxtrnewsymbol[description={complex species}]{complspec}{\ensuremath{C}}
\def\complspec{{\ensuremath{\gls{complspec}}}}

\glsxtrnewsymbol[description={strand length}]{strandlength}{\ensuremath{L}}
\def\strandlength{{\ensuremath{\gls{strandlength}}}}

\glsxtrnewsymbol[description={mean strand length}]{meanstrandlength}{\ensuremath{\overline \strandlength}}

\glsxtrnewsymbol[description={hybridization Gibbs free energy}]{Ghyb}{\ensuremath{\Delta \gibbs_\text{hyb}}}
\def\Ghyb{{\ensuremath{\gls{Ghyb}}}}

\glsxtrnewsymbol[description={total free energy of a hybridized complex}]{Gtot}{\ensuremath{\Delta \gibbs_\text{tot}}}

\glsxtrnewsymbol[description={hybridization rate constant}]{hrc}{\ensuremath{k_\text{on}}}

\glsxtrnewsymbol[description={dehybridization rate constant}]{drc}{\ensuremath{k_\text{off}}}

\glsxtrnewsymbol[description={symmetry factor}]{symfac}{\ensuremath{\rho}}

\glsxtrnewsymbol[description={number of possible hybridization channels}]{hcn}{\ensuremath{\chi}}

\glsxtrnewsymbol[description={dissociation constant}]{discon}{\ensuremath{K_\mathrm{D}}}
\def\discon{{\ensuremath{\gls{discon}}}}

\glsxtrnewsymbol[description={effective dissociation constant}]{mdiscon}{\ensuremath{\tilde K_\mathrm{D}}}
\def\mdiscon{{\ensuremath{\gls{mdiscon}}}}

\def\seds{{\ensuremath{\gamma}}}

\glsxtrnewsymbol[description={average energy value of complementary (nearest neighbor) blocks}]{aevocb}{\ensuremath{\overline \seds_\text{com}}}
\def\aevocb{{\ensuremath{\gls{aevocb}}}}

\glsxtrnewsymbol[description={stacking energy for blocks with one non-complementary nucleotide pair}]{seImm}{\ensuremath{\seds_\text{1nc}}} %stacking energy one missmatch
\def\seImm{{\ensuremath{\gls{seImm}}}}

\glsxtrnewsymbol[description={stacking energy for blocks with two non-complementary nucleotide pairs}]{seIImm}{\ensuremath{\seds_\text{2nc}}} %stacking energy two missmatches
\def\seIImm{{\ensuremath{\gls{seIImm}}}}

\glsxtrnewsymbol[description={stacking energy difference (hybridization bias) of alternating and homogeneous blocks}]{sedoahb}{\ensuremath{\Delta \seds}}

\glsxtrnewsymbol[description={lower bound dehybridization length}]{lbdl}{\ensuremath{l_\text{low}}}% (for completely complementary hybridization with more nucleotides will dehybridize with the same rate)

\def\sede{\ensuremath{\epsilon}}
\glsxtrnewsymbol[description={average energy value of complementary blocks with dangling end}]{aevocbde}{\ensuremath{\overline \sede_\text{com}}}
\def\aevocbde{{\ensuremath{\gls{aevocbde}}}}

\glsxtrnewsymbol[description={stacking energy for blocks with dangling end and one non-complementary nucleotide pair}]{sedeImm}{\ensuremath{\sede_\text{1nc}}} %stacking energy one missmatch
\def\sedeImm{{\ensuremath{\gls{sedeImm}}}}

\glsxtrnewsymbol[description={stacking energy difference of alternating and homogeneous blocks with dangling end}]{sededoahb}{\ensuremath{\delta_\sede}}

\glsxtrnewsymbol[description={ligation rate constant}]{lrc}{\ensuremath{k_\text{lig}}}
\def\lrc{{\ensuremath{\gls{lrc}}}}
\glsxtrnewsymbol[description={default ligation rate constant}]{plrc}{\ensuremath{\lrc_0}}
\def\plrc{{\ensuremath{\gls{plrc}}}}

\glsxtrnewsymbol[description={ligation length}]{liglen}{\ensuremath{l_\text{lig}}}

\def\stallingfactorsymbol{\ensuremath{\Phi}}
\glsxtrnewsymbol[description={stalling factors}]{stalfac}{\ensuremath{\stallingfactorsymbol_\pm}}
\def\stalfac{{\ensuremath{\gls{stalfac}}}}
\def\stalfacp{\ensuremath{\stallingfactorsymbol_+}}
\def\stalfacm{\ensuremath{\stallingfactorsymbol_-}}

\glsxtrnewsymbol[description={complementarity, index $\pm i, i \in \{ 1,2\}$ indicates relative position to the ligation spot: $\pm$ left/right, nearest ($1$) or nextnearest ($2$) neighbour}]{complementarities}{\ensuremath{\kappa}}

\def\stalparsym{\ensuremath{\sigma}}
\glsxtrnewsymbol[description={stalling parameter, $i \in \{1,2\}$}]{stalpars}{\ensuremath{\stalparsym_i}}
\def\stalparI{{\ensuremath{\stalparsym_1}}}
\def\stalparII{{\ensuremath{\stalparsym_2}}}

\def\cleavagesym{{\ensuremath{k_\text{cut}}}}
\glsxtrnewsymbol[description={cleavage rate constant}]{clrc}{\ensuremath{\cleavagesym}}

\glsxtrnewsymbol[description={effective cleavage rate constant}]{ecrc}{\ensuremath{\tilde k_\text{cut}}}

\glsxtrnewsymbol[description={collision time scale}]{t0}{\ensuremath{\timevar_0}}

\glsxtrnewsymbol[description={collision rate constant }]{crc}{\ensuremath{k_\text{coll}}}

\glsxtrnewsymbol[description={motif length}]{motiflength}{\ensuremath{\ell}}
\def\motiflength{{\ensuremath{\gls{motiflength}}}}

\glsxtrnewsymbol[description={motif space}]{motifspace}{\ensuremath{\mathcal{M}_\motiflength}}

\glsxtrnewsymbol[description={total number of motifs}]{totalnumberofmotifs}{\ensuremath{\mathcal{N}}}

\glsxtrnewsymbol[description={total number of strands}]{totalnumberofstrands}{\ensuremath{\mathcal{N}_\text{strands}}}

\glsxtrnewsymbol[description={extension rate constant}]{eerc}{\ensuremath{k_\text{ext}}}

\glsxtrnewsymbol[description={motif production rate constants}]{mprc}{\ensuremath{\tilde k_\text{ext}}}
\def\mprc{{\ensuremath{\gls{mprc}}}}

\glsxtrnewsymbol[description={motif production rate constants standard deviation}]{mprcstd}{\ensuremath{\sigma(\mprc)}}

\glsxtrnewsymbol[description={motif production exposure}]{E}{\ensuremath{\mathcal{E}}}
\def\mpe{{\ensuremath{\gls{E}}}}

\glsxtrnewsymbol[description={motif production rate}]{M}{\ensuremath{\hat \Lambda}}
\def\mpr{{\ensuremath{\gls{M}}}}

\glsxtrnewsymbol[description={expected motif production counts}]{eM}{\ensuremath{\Lambda}}
\def\empc{{\ensuremath{\gls{eM}}}}

\glsxtrnewsymbol[description={breakage rate constant}]{brc}{\ensuremath{\tilde k_\text{cut}}}
\def\brc{{\ensuremath{\gls{brc}}}}

\newcommand\mbr[2]{\ensuremath{\Gamma^{(#1)}_{(#2)}}}
\glsxtrnewsymbol[description={motif breakage rate}]{mbr}{\ensuremath{\mbr{i}{\motifvar}}}

\glsxtrnewsymbol[description={effective breakage rate constant}]{ebf}{\ensuremath{\brc}}%{\ensuremath{{\tilde k^{(i)}_\text{cut}}_{\motifvar}}}

\glsxtrnewsymbol[description={breakage rate}]{brt}{\ensuremath{\beta}}

\glsxtrnewsymbol[description={influx rate constant}]{irc}{\ensuremath{\iota}}

\glsxtrnewsymbol[description={outflux rate constant}]{orc}{\ensuremath{\omega}}

\glsxtrnewsymbol[description={average motif production rate (Poisson rate)}]{ampr}{\ensuremath{\tilde{ \lambda}}}

\glsxtrnewsymbol[description={motif ode terms}]{mot}{\ensuremath{\vek{f}}}
\def\mot{{\ensuremath{\gls{mot}}}}

\glsxtrnewsymbol[description={influx ode terms}]{iot}{\ensuremath{\mot_\text{in}}}

\glsxtrnewsymbol[description={outflux ode terms}]{oot}{\ensuremath{\mot_\text{out}}}

\glsxtrnewsymbol[description={motif production ode terms}]{mpot}{\ensuremath{\mot_\text{lig}}}

\glsxtrnewsymbol[description={breakage ode terms}]{bot}{\ensuremath{\mot_\text{break}}}

\glsxtrnewsymbol[description={motif rate terms}]{mrt}{\ensuremath{\mot_\text{lig,break}}}

\glsxtrnewsymbol[description={pseudo count}]{pseudoI}{\ensuremath{\delta_1}}

\glsxtrnewsymbol[description={pseudo zero}]{pseudoO}{\ensuremath{\delta_0}}

\glsxtrnewsymbol[description={smooth clipping}]{smcl}{\ensuremath{\widetilde \Theta}}

\glsxtrnewsymbol[description={strand completion rate constant}]{scrc}{\ensuremath{k_\text{strand}}}

\glsxtrnewsymbol[description={mass conservation rate constant}]{mcrc}{\ensuremath{k_\text{mass}}}

\glsxtrnewsymbol[description={mass conservation rate}]{mcr}{\ensuremath{\mot_\text{mass}}}

\glsxtrnewsymbol[description={mass conservation rate potential}]{mcrp}{\ensuremath{U_\text{mass}}}

\glsxtrnewsymbol[description={infinitely long strands concentration}]{ilsc}{\ensuremath{\c_\infty}}

\glsxtrnewsymbol[description={acceleration of motif production rate constants}]{amprc}{\ensuremath{\varphi}}

\glsxtrnewsymbol[description={dangling motif production rate constants}]{dmprc}{\ensuremath{\zeta}}

\glsxtrnewsymbol[description={sequence dependent motif production rate constant}]{smprc}{\ensuremath{\varsigma}}

\glsxtrnewsymbol[description={temperature dependence of the motif production rate constants}]{tmprc}{\ensuremath{\tau}}

\def\p{\ensuremath{p}}
\glsxtrnewsymbol[description={left produced motif}]{lpm}{\ensuremath{\p_\mathrm{l}}}

\glsxtrnewsymbol[description={central produced motif}]{cpm}{\ensuremath{\p_\mathrm{c}}}
\def\cpm{{\ensuremath{\gls{cpm}}}}

\glsxtrnewsymbol[description={right produced motif}]{rpm}{\ensuremath{\p_\mathrm{r}}}

\glsxtrnewsymbol[description={product template motif}]{tpm}{\ensuremath{ t }}
\def\tpm{{\ensuremath{\gls{tpm}}}}

\glsxtrnewsymbol[description={left reactant motif}]{lrm}{\ensuremath{ l }}
\def\lrm{{\ensuremath{\gls{lrm}}}}

\glsxtrnewsymbol[description={right reactant motif}]{rrm}{\ensuremath{ r }}
\def\rrm{{\ensuremath{\gls{rrm}}}}

\glsxtrnewsymbol[description={ode terms for left reactant}]{lrmot}{\ensuremath{\mot_{\text{lig}_\mathrm{l}}}}

\glsxtrnewsymbol[description={ode terms for right reactant}]{rrmot}{\ensuremath{\mot_{\text{lig}_\mathrm{r}}}}

\glsxtrnewsymbol[description={ode terms for produced motifs}]{pmot}{\ensuremath{\mot_{\text{lig}_\mathrm{p}}}}

\def\confreq{\ensuremath{\phi}}

\glsxtrnewsymbol[description={left continuation frequency}]{lconfreq}{\ensuremath{ \confreq_- }}

\glsxtrnewsymbol[description={right continuation frequency}]{rconfreq}{\ensuremath{ \confreq_+ }}

\def\0{0}%\emptyset}
\def\reprod{\ensuremath{{\cpm,\tpm}}}
\def\reproduction{\ensuremath{{\cpm(\lrm,\rrm),\tpm}}}
\def\reactants{{\ensuremath{ {\lrm,\rrm,\tpm} }}}

\newcommand\ex[1]{{\ensuremath{e^{#1}}}}

\glsxtrnewsymbol[description={strandsampler weights}]{ssw}{\ensuremath{ \hat \mc }}

\glsxtrnewsymbol[description={signal}]{signal}{\ensuremath{ s }}

\glsxtrnewsymbol[description={data (ligation counts)}]{data}{\ensuremath{ d }}
\def\data{{\ensuremath{\gls{data}}}}

\glsxtrnewsymbol[description={entropy}]{entropy}{\ensuremath{ S }}

\glsxtrnewsymbol[description={system level zebraness: average fraction of alternating vs. homogeneous binary motis over all strands weighted by their length.}]{slz}{\ensuremath{ Z }}
\def\slz{{\ensuremath{\gls{slz}}}}

\makenoidxglossaries

\usepackage{pgf}

\newcommand\ignoriere[1]{\iffalse #1 \fi}
\usepackage{algorithm}
\usepackage[noend]{algpseudocode}

\newcommand\mpaadress{\href{https://ror.org/017qcv467}{Max-Planck-Institut f{\"u}r Astrophysik},
Information Field Theory Group,
Karl-Schwarzschild-Str.~1,
85748~Garching,
Germany}
\newcommand\lmuadress{\href{https://ror.org/05591te55}{Ludwig-Maximilians-Universit{\"a}t M{\"u}nchen},
Fakult{\"a}t f{\"u}r Physik,
Geschwister-Scholl-Platz~1,
80539~Munich,
Germany}
\newcommand\qbiotumadress{\href{https://ror.org/02kkvpp62}{Technical University of Munich},
TUM School of Natural Sciences,
Department of Bioscience,
James-Franck-Str.~1,
85748~Garching,
Germany}
\newcommand\originsadress{\href{https://ror.org/010wkny21}{Exzellenzcluster ORIGINS},
Boltzmannstr.~2,
85748~Garching,
Germany}
\newcommand\dzaadress{Deutsches Zentrum für Astrophysik,
  Postplatz~1,
  02826~G{\"o}rlitz,
  Germany}

\newcommand\old[1]{}
\usepackage{authblk}

\author[1,3,4,5,6]{Johannes~Harth-Kitzerow}
\author[4,5,7]{Ulrich~Gerland}
\author[1,2,3,5,8]{Torsten~A.~En{\ss}lin}
\affil[1]{\small{\mpaadress}}
\affil[2]{\small{\dzaadress}}
\affil[3]{\small{\lmuadress}}
\affil[4]{\small{\qbiotumadress}}
\affil[5]{\small{\originsadress}}
\affil[6]{\small{jharthki@mpa-garching.mpg.de}}
\affil[7]{\small{gerland@tum.de}}
\affil[8]{\small{ensslin@mpa-garching.mpg.de}}

\usepackage{cite}

\begin{document}

\title{Bayesian Rate Inference for Sequence Motif Dynamics in Systems of Reactive Nucleic Acids}

\makeatletter
\renewcommand\@fnsymbol[1]{\@arabic{#1}\ }
\makeatother
\maketitle

\section*{Abstract}
\hrule
~\linebreak
The RNA world hypothesis suggests a pathway of how life emerged on
early earth.
It assumes that life started with RNA based systems, capable of storing, transmitting and
replicating information,
envisioning that
monomers and short RNA oligomers interact to form longer strands,
eventually becoming catalytically active ribozymes.
Key reactions in RNA pools are hybridization, dehybridization, templated ligation, and cleavage.
Those reactions depend on many environmental parameters and the wide range of possible configurations
among interacting strands.
In order to scan such high dimensional parameter spaces, efficient descriptions
are needed.
Motif rate equations project complex strand reactor dynamics onto sequence
motif space.
Here we present a Bayesian inference framework to infer their parameters from
ligation count data produced by strand reactor simulations.
This provides a framework to match the simpler motif rate equations to more
complex simulations.
Additionally, it is a step towards inferring reaction rate constants
directly from experimental data, including rigorous uncertainty estimation.
This could be an essential procedure to connect theory and experiment,
and deepen our understanding of the essential features necessary for life to emerge.

\linebreak%\vspace{\textheight}
\hrule

\section{Introduction}

One research line within the origins of life field is based on the RNA world hypothesis,
which posits that life can emerge from inter-reacting RNA or other
polynucleotide
strands\cite{Crick1968TheOrigin,Orgel1968Evolution,Szostak2012TheEightfold,Higgs2015TheRNAWorld,Pressman2015TheRNAWorld}. 
These strands, which are made up from nucleotides such as A, U, C, and G for RNA, would store, transmit, and replicate information.
They interact by hybridizing to each other, forming complexes, which enable reactions such as templated ligation and
cleavage (hydrolysis) of strands.
In templated ligation, an RNA template catalyzes the ligation
of two partially complementary strands, which are adjacently bound and eventually ligate to form a single, longer strand.
This effectively enables information transmission and, thus, 
the survival of information longer than the life time of individual strands.

Longer strands of RNA can fold into catalytically active ribozymes, which could accelerate the replication of their own or other sequences~\cite{Zaug1986Intervening,FerreDAmare2010Small,Higgs2015TheRNAWorld}.
This could lead to the formation of cooperative reaction networks that enable sequence replication with genetic 
heredity and mutation, and thus Darwinian evolution~\cite{Higgs2015TheRNAWorld}. 
Recently, polymerase ribozymes were discovered that only contain 45 nucleotides~\cite{Gianni2026ASMall}.

Despite that we can build on extensive prior
work~\cite{Obermayer2011Emergence,Derr2012Prebiotically,Vaidya2013Recycling,Tkachenko2015Spontaneous,Fellermann2017Sequence,Tupper2017Role,Matsubara2018Kinetic,Tkachenko2018Onset,Toyabe2019Cooperative,Rosenberger2021SelfAssembly,Tupper2021Rolling,Goeppel2022Thermodynamic,Laurent2024Emergence,Tkachenko2024Emergence,HarthKitzerow2026Sequence},
achieving a
quantitative understanding of interacting
polynucleotide molecules remains a major challenge.
Experimentally, such assemblies are accessible through techniques such as mass
spectroscopy~\cite{Levy1998TheStability} or sequencing~\cite{Serrao2024Replication}. 
However, many relevant candidate systems exist, consisting of RNA, DNA, or 
other nucleotides~\cite{Taylor2015Catalysts}.
Also, the number of different nucleotides is uncertain.
While prebiotic polynucleotides may have consisted of more than four different nucleotides, 
binary alphabets are also considered~\cite{Crick1968TheOrigin,Orgel1968Evolution,Levy1998TheStability,Haenle2018EnzymeFree}.
In addition, many parameters like pH, ion concentrations, and temperature strongly affect the
kinetics of nucleic acid reactions.
To scan such large parameter spaces and infer reaction rate constants from
experimental data, effective simulations and inference algorithms are needed.

Strand reactor simulations based on Gillespie's stochastic simulation algorithm
give insight on the emergence of characteristic length distributions and
symmetry breaking in sequence space~\cite{Goeppel2022Thermodynamic,Rosenberger2021SelfAssembly},
but are computationally highly expensive and,
thus, infeasible for scanning large parameter spaces.
To reduce complexity, one can scan strands for all containing sequence motifs,
\ie\ subsequences up to a certain length $\motiflength$
equal to the maximum number of nucleotides in a motif.
Reduced models use this approach and integrate more easily tractable ordinary differential
equations~\cite{Tkachenko2015Spontaneous,Tkachenko2018Onset}, 
but are not directly connected to experimental parameters.
In an earlier work, we developed chemical rate equations for sequence motifs,
called motif rate equations,
which are directly connected to strand reactor simulations and their parameters~\cite{HarthKitzerow2026Sequence}.
%TODO: further in silico simulation:~\cite{Wu2012TheOrigin}

Here we present a Bayesian rate inference method to infer
motif extension rate constants from templated ligation counts,
\ie\ the number of templated ligation reactions at each time step.
Those motif extension rate constants are effective ligation rate constants for
sequence motifs that consider only the nucleotides directly at the ligation spot
and incorporate hybridization and dehybridization implicitly.
Also, they serve as parameters for motif rate equations.
For their inference, we set up a Bayesian data model.
It consists of a hyperprior, a prior and the likelihood
that hierarchically connect the strand reactor parameters with the data.
As hyperprior, we use the motif extension rate constants model from
Ref.~\cite{HarthKitzerow2026Sequence} that computes
the motif extension rate constants from the strand reactor parameters.
This hyperprior determines the parameters of a log-normal prior
for the motif extension rate constants.
As part of the likelihood, the motif extension rate constants
are transformed to expected ligation counts, which are directly connected to the
data.
With the help of prior and likelihood functions, a posteroir probability is
defined.
We draw samples from this using geometric Variational Inference (geoVI)~\cite{Frank2021GeoVI}.

The ligation counts that serve as data for the inference are produced by
RNA strand reactor simulations~\cite{Goeppel2022Thermodynamic}.
For that we mimic typical scenarios,
where one knows the parameters roughly, but not exactly.
Specifically, we use prior values
that are uninformative with respect to the reactants' sequences and
ligation stalling in the presence of mismatches at the ligation site.
For the simulated data, we choose two scenarios with different parameters,
first with ligation stalling only and,
second, the same but with hybridization bias favouring alternating sequences to bind.

This framework has two types of application.
One being the first step of building a Bayesian analysis framework
for nucleic acid reactors, to link theoretical models to experimental data.
The other being a tool to improve the projection of strand reactor
simulations on motif rate equations as so far, only approximately estimated 
reaction rate constants have been computed that neglect effects like longer hybridization
sites.
For simplicity, in this paper, we assume a binary alphabet, $\alphabet$,
and call our nucleotides $\letterI$ and $\letterII$. 

In the following, we first summarize the Bayesian data model used,
starting with the likelihood deriving expected ligation counts
from motif extension rate constants in Section~\ref{sec:Likelihood}.
In Section~\ref{sec:Prior}, we continue with the summary of the hybridization and templated ligation models.
We proceed in Section~\ref{sec:Evidence} with the choice of the parameter regimes for the strand reactor
simulations providing the data,
to conclude the methods section with a short summary of geoVI (Section~\ref{sec:GeoVI}). 
Finally, we apply the inference framework on simulated data in
Section~\ref{sec:results},
first, in Section~\ref{sec:Stalling}, without and then,
in Section~\ref{sec:HybBias},
with hybridization energy bias on alternating sequences.

\section{Methods}

\begin{figure}
    \centering
    \includegraphics[width=\columnwidth]{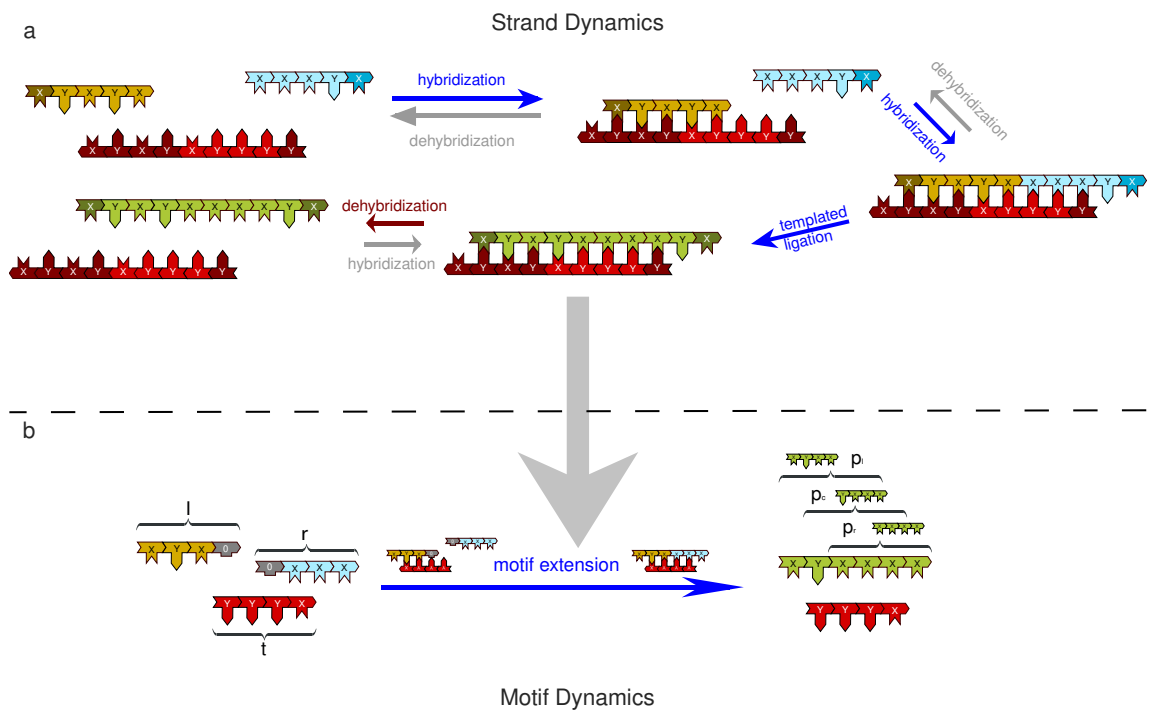}
    \caption{\label{fig:motif_production_strandBRI}
    (a) Simulation model of the strand reactor and (b) its projection onto motif rate equation.
    }
\end{figure}

The aim of this section is to find reaction rate constants
in a nucleic acid reactor that performs templated ligation,
hybridization and dehybridization (Fig.~\ref{fig:motif_production_strandBRI}),
given templated ligation reaction counts.
Templated ligation kinetics are described by templated ligation rate constants and
the concentrations of the reactants.
We describe the reaction dynamics in sequence motif space~\cite{HarthKitzerow2026Sequence,Tkachenko2015Spontaneous,Tkachenko2018Onset}.
There, exact hybridization configurations are not tracked.
Instead, we estimate the fraction of hybridized configurations
from corresponding hybridization energies
and combine them with the template ligation rate constants, $\lrc$,
to obtain motif extension rate constants, $\mprc$
that contain the hybridization-dynamics implicitly.

Inferring motif extension rate constants $\mprc$ from ligation counts $\data$
is an inverse problem.
Given the initial concentrations in the nucleic acid pool,
the ligation counts are a consequence of sequence motifs reacting with each
other given their current concentration and the motif extension rate constants.
To infer those from the ligation counts, we use Bayes' theorem:
\begin{align}
    \condP{\mprc}{\data} = \frac{
        \condP{\data}{\mprc} \Prob{\mprc}
    }{
        \Prob{\data}
    }
\end{align}
Bayes' theorem connects four probabilities:
First, the likelihood $\condP{\data}{\mprc}$
that expresses the probability of the observed data to have been generated
given assumed motif extension rate constants, $\mprc$.
These rate constants determine
the expected number of ligation counts
via the motif rate equations.
The observed data is assumed to be drawn from a Poisson distribution with
these expected counts.
Second, the prior $\Prob{\mprc}$,
which includes our a priori knowledge of the motif extension rate constants.
Third, the probability of the data irrespective of the unknown rate constants, $\Prob{\data}$, called evidence,
which can be understood as the normalization constant of the fourth
probability, the posterior, with respect to the rate constants.
In our case, the ligation counts serve as the data,
which we obtain from more expensive simulations,
whose parameters will be discussed in this context.
The posterior, $\condP{\mprc}{\data}$,
can be computed numerically from the other probabilities.
In the following, we will go over those parts one after the other.

\subsection{Likelihood: Motif Rate Equations with Poisson Noise}\label{sec:Likelihood}

For the sake of this paper, we assume that the concentrations of the sequence motifs
are known at all times as these are directly accessible from the simulations.
%\begin{align}
%    \Prob{\c(\timevar)} = \delta\left(\c(\timevar) - \c_\timevar \right)
%\end{align}
Let $\mcv\left(\timevar\right)$ be the motif concentration vector storing the concentration of each
motif at time $\timevar$.
In one motif extension reaction, a left (ending) sequence motif $\lrm$ is
extended by a right (beginning) sequence motif $\rrm$ that are both hybridized
adjacently onto a template motif $\tpm$ with the ligation site at its center.
In the following, we refer to the combination of the two reactants and the
template as a single reaction channel.
Since we track sequence motifs of a certain length ($\motiflength$)
(and strands that are two nucleotides shorter than that length explicitly),
the resulting extended strand might contain more nucleotides than one sequence motif.
We thus have to scan it again with respect to its containing sequence motifs.
In addition, the template strand is also only captured as its sequence motif
centered directly at the ligation site.
Consequently, only the hybridization partner of the resulting sequence
motif with the ligation site at its center is fully known.
For a given time interval $[0,T]$, the integrated motif extension rate of reactants
$\lrm$ and $\rrm$ with template $\tpm$ building such a central sequence motif $\cpm(\lrm,\rrm)$ is then
\begin{align}
    \mpr_\reactants &= \int_0^T \mprc_\reactants \c_\lrm \c_\rrm
    \c_\tpm \mathrm d \timevar.
\end{align}
In a reactor with a volume $\vol$, we thus expect
$\empc_\reactants = \vol \mpr_\reactants$ motif extension counts.
Multiple reaction channels produce the same central sequence motif.
Tracking the dependence of the integrated extension rate
on the central produced motif that aligns perfectly with the template gives,
\begin{align}
    \mpr_\reprod &= \sum_{\lrm,\rrm \vert \cpm(\lrm,\rrm)=\cpm} \int_0^T \underbrace{\mprc_\reactants}_{\equiv\mprc_\reproduction} \c_\lrm \c_\rrm
    \c_\tpm \mathrm d \timevar.
\end{align}
For time independent reaction rate constants, we can pull the rate constants
out of the time integral.
\begin{align}\label{eq:rrc}
    \mpr_\reprod &= \mprc_\reprod \mpe_\reprod,
\end{align}
with reaction channel exposure,
\begin{align}
    \mpe_\reprod\left[ \mcv \right] := \sum_{\lrm,\rrm \vert \cpm(\lrm,\rrm)=\cpm} \int_0^T \c_\lrm \c_\rrm \c_\tpm \mathrm d \timevar.
\end{align}
The motif extension rates are the expected ligation counts specific to the
corresponding reaction channel.
Assuming shot noise, we get a Poisson likelihood after marginalizing
over the reaction rates $\mpr_\reactants$.
\begin{align}
%    \condP{\data}{\mprc} &= \int \condP{\data}{\mprc,\c\left(\timevar\right)}
%    \underbrace{\Prob{\c\left(\timevar\right)}}_{\delta\left(\c\left(\timevar\right)-\c_\timevar\right)}
%    \mathrm d \c\left(\timevar\right).
%    \\
    &\condP{\data_\reprod}{\mprc_\reprod,\mcv(\timevar)}\notag
    \\
    = &\int \condP{\data_\reprod}{\empc_\reprod}
    \delta\left(\empc_\reprod -\vol \mprc_\reprod\mpe_\reprod\left[\mcv\left(\timevar\right)\right]\right)
    \mathrm d \empc_\reprod
    \notag \\
    = &\left. \Poissonian{\data_\reprod}{\empc_\reprod}
    \right\vert_{\empc_\reprod=\vol \mprc_\reprod\mpe_\reprod\left[\mcv\left(\timevar\right)\right]}.
\end{align}
Due to statistical independence of the ligation counts of different reaction channels,
the total likelihood that takes all reaction channels into account,
is the product of the individual ones.
\begin{align}
    \condP{\data}{\mprc,\mcv(\timevar)}
    &= \prod_\reprod
    \condP{\data_\reprod}{\mprc_\reprod,\mcv(\timevar)}.
\end{align}

\subsection{A Priori Motif Extension Rate Constants}\label{sec:Prior}

\begin{figure}
	\centering
	\begin{tikzpicture}[
		roundnode1/.style={circle, draw=black!60, fill=white!5, very thick, minimum size=3mm},
		roundnode2/.style={circle, dashed, draw=black!60, fill=white!5, very thick, minimum size=3mm},
		rectnode1/.style={rectangle, draw=black!60, fill=white!5, very thick, minimum size=3mm},
		rectnode2/.style={rectangle, rounded corners=3pt, draw=black!60, fill=white!5, very thick, minimum size=3mm},
	]
	%Nodes
	\coordinate (aa);
	\node[roundnode2] (plrcnode) [left of=aa] {$\plrc$};
	\node[roundnode2] (stalparsnode) [right of=aa] {$\stalparsym$};
	\node[roundnode2] (lrcnode) [below of= aa]{$\lrc$};
	\draw[->] (plrcnode.south east) -- (lrcnode.north west);
	\draw[->] (stalparsnode.south west) -- (lrcnode.north east);

	\coordinate (ab) at (3,0);% ($(sedenode.east)!0.5!(sedsnode.west)$);
	\node[roundnode2] (sedenode) [left of=ab]{$\sede$};
	\node[roundnode2] (sedsnode) [right of=ab]{$\seds$};
	\node[roundnode2] (mdisconnode) [below of= ab]{$\mdiscon$};
	\draw[->] (sedenode.south east) -- (mdisconnode.north west);
	\draw[->] (sedsnode.south west) -- (mdisconnode.north east);

	\coordinate (ac) at ($(lrcnode.east)!0.5!(mdisconnode.west)$);

	\node[roundnode1] (mprcnode) [below of= ac]{$\mprc$};
	\node[rectnode2] (mctnode) [right of=mdisconnode, below of=mdisconnode] {$\mct$};
	\coordinate (a) at ($(mprcnode.east)!0.5!(mctnode.west)$);
	\draw[->] (lrcnode.south east) -- (mprcnode.north west);
	\draw[->] (mdisconnode.south west) -- (mprcnode.north east);

	\node[roundnode1]  (mprnode) [below of= a] {$\empc$};
	\node[rectnode1] (datanode) [below of= mprnode]{$\data$};
	%Lines
	\draw[->] (mctnode.south west) -- (mprnode.north east);
	\draw[->] (mprcnode.south east) -- (mprnode.north west); %.. controls +(down:10mm) and +(left:1mm) .. (LambdaNode.north);
	\draw[->] (mprnode.south) -- (datanode.north);
\end{tikzpicture}
    \caption{\label{fig:model}Inference model:
    Prior parameters (dashed circle),
    inferred parameters (solid circles),
    likelihood parameters (round cornered rectangle) and
    given values (rectangle).}
\end{figure}
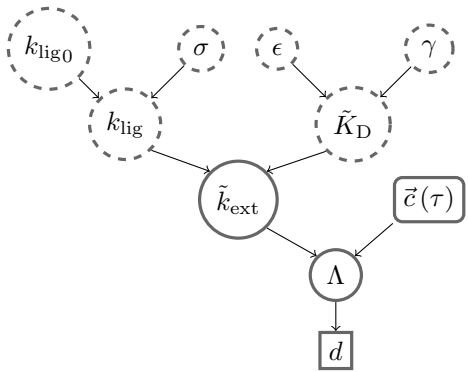

To estimate the motif extension rate constants a priori,
we consider the hybridization energy and templated ligation rate constants
model of Ref.~\cite{Goeppel2022Thermodynamic}
adjusted onto sequence motif space as in Ref.~\cite{HarthKitzerow2026Sequence}
(Fig.~\ref{fig:model}).
There, the motif extension rate constants are computed as the fraction of
templated ligation rate constants and dissociation constants.
\begin{align}\label{eq:MPRC}
    \mprc = \frac{\lrc}{\discon}.
\end{align}
The templated ligation rate constants are explicitly set by a constant,
$\plrc$, and stalling parameters $\stalparsym=\left(\stalparI,\stalparII\right)$
that define stalling in the presence of mismatches directly at the ligation
site.
\begin{align}\label{eq:Stalling}
    \lrc = \plrc \stalfacp \stalfacm,
\end{align}
with stalling factors
\begin{align}
    \stalfac = &\begin{cases}
        \stalparI \stalparII &\text{two mismatches${^\star}$,}
        \\
        \stalparI&\text{one mismatch directly${^\star}$,}
        \\
        \stalparII&\text{one mismatch neighbouring${^\star}$,}
        \\
        1&\text{no mismatches${^\star}$.}
    \end{cases}
        \\
        &\quad \text{${}^\star$up-/downstream of the ligation site}\notag
\end{align}
The dissociation constant is given by the Gibbs factor, which is the
exponentiated Gibbs free energy difference between the hybridized and the
dehybridized state of two (or more) strands forming a complex $\complspec$.
\begin{align}
    \discon = \ex{\Ghyb \left( \complspec \right)}.
\end{align}
The Gibbs free energy is computed per block of two nucleotides width
and then summed up.
If block $i$ has a dangling end, it contributes $\sede_i$ to the hybridization energy,
if it is a blunt end or continuous two by two nucleotide block, it contributes $\seds_i$.
Depending on the number of mismatches inside a block, and whether containing an
alternating or homogeneous dimer (hybridization bias), those contributions can have different predefined values.
\begin{align}
    \Ghyb\left( \complspec \right) = \sum_{i \text{ block in } \complspec} \sede_i + \seds_i.
\end{align}

To validate our method, we emulate a scenario,
where one knows the parameters roughly but not exactly.
Those rough values determine the mean of a multivariate Gaussian prior
with a diagonal covariance matrix $\PriorCov{\mprc}$,
\begin{align}
    \Prob{\mprc}&=\Gauss{\mprc-\ExpV{\mprc}}{\PriorCov{\mprc}}\text{, with}
    \\
    \Gauss{x}{X} &:=\Gaussian{x}{X}.
\end{align}
The covariance is chosen to allow a standard deviation of one order of
magnitude per time unit $\timeunit$,
which is $10^{-4} \frac{\mathrm{L}^2}{\mathrm{mol}^2\timeunit}$
for our parameters that we discuss in the next section.

On the computational side, we ensure positivity of the reaction rate constants by 
tracking their logarithm such that the final prior is a Log-normal distribution.
For the computation, samples are drawn from a standard Gaussian and linearly transformed
such that their exponentiated mean fits our expected motif extension rate constants.
Mathematically, this is equivalent to
pulling the linear transformation into the likelihood
such that the prior follows a standard Gaussian and the likelihood
a Log-normal Poisson distribution~\cite{Knollmueller2018Encoding}.

\subsection{Evidence: Strand Reactor Simulation Data}\label{sec:Evidence}

Specifically, we use parameter set~0 in Ref.~\cite{HarthKitzerow2026Sequence}
as prior mean to infer the parameters of sets~1 (scenario~1) and~3 (scenario~2)
from data simulated by the strand reactor simulation of Ref.~\cite{Goeppel2022Thermodynamic}
with corresponding parameters.
These parameter sets are distinguished by different values for ligation
stalling and hybridization bias.
While parameter set~0 does not stall ligation ($\stalparI=\stalparII=0$),
parameter sets~1 and 3 do ($\stalparI=1,\stalparII=0.05$).
In addition, parameter set~3 introduces a hybridization bias of $-0.3$ for
$2\times 2$-blocks and $-0.15$ for dangling blocks
such that alternating blocks without a mismatch bond stronger than homogeneous
blocks.
The other two parameter sets do not discriminate alternating and homogeneous
blocks.
For all parameter sets, the mean for hybridization energy contributions and the
ligation rate constant without stalling are the same,
see Table~\ref{tab:HybEnergyParameters}.
Cleavage rate constants are chosen negligible small.
\begin{table}
    \centering
	\begin{tabular}{p{7em} cccccc}
	    Parameter & \plrc & $\aevocbde$& $\sedeImm$ & $\aevocb$ & $\seImm$ & $\seIImm$ \\
	    \hline
	    Value & $3.73\cdot 10^{-6}$ & $-0.625$ & $0.375$ & $-1.25$ & $0.375$ & $0.75$ \\
    \end{tabular}
    \caption{\label{tab:HybEnergyParameters}
        Reaction parameters of the strand reactor simulation.
        The indices ``1nc'' and ``2nc'' stand for one and two non-complementary mismatches,
        ``com'' for complementary, i.\,e.\ without mismatches.
    }
\end{table}

\subsection{Posterior Estimation with Geometric Variational Inference}\label{sec:GeoVI}

To estimate the posterior we use geometric Variational Inference (geoVI)~\cite{Frank2021GeoVI},
an improvement of Metric Gaussian Variational Inference (MGVI)~\cite{Knollmueller2020MGVI}.
MGVI approximates a complex posterior distribution by a Gaussian by minimizing
the Kullback-Leibler divergence.
Alternately,
posterior mean and posterior covariance are estimated from
samples drawn from the
current approximation of the posterior
using the inverse Fisher information metric as approximation of the likelihood covariance
and the Wiener Filter to compute the posterior covariance from that~\cite{Knollmueller2020MGVI}.
Additionally, geoVI determines a coordinate transformation for which the
posterior is most similar to a standard Gaussian distribution.
This way, one can approximate complex distributions shaped very differently
to a Gaussian~\cite{Frank2021GeoVI}.

\section{Results and Discussion}\label{sec:results}

\subsection{Ligation Stalling}\label{sec:Stalling}

\begin{figure}
    \centering
    \includegraphics[width=\columnwidth]{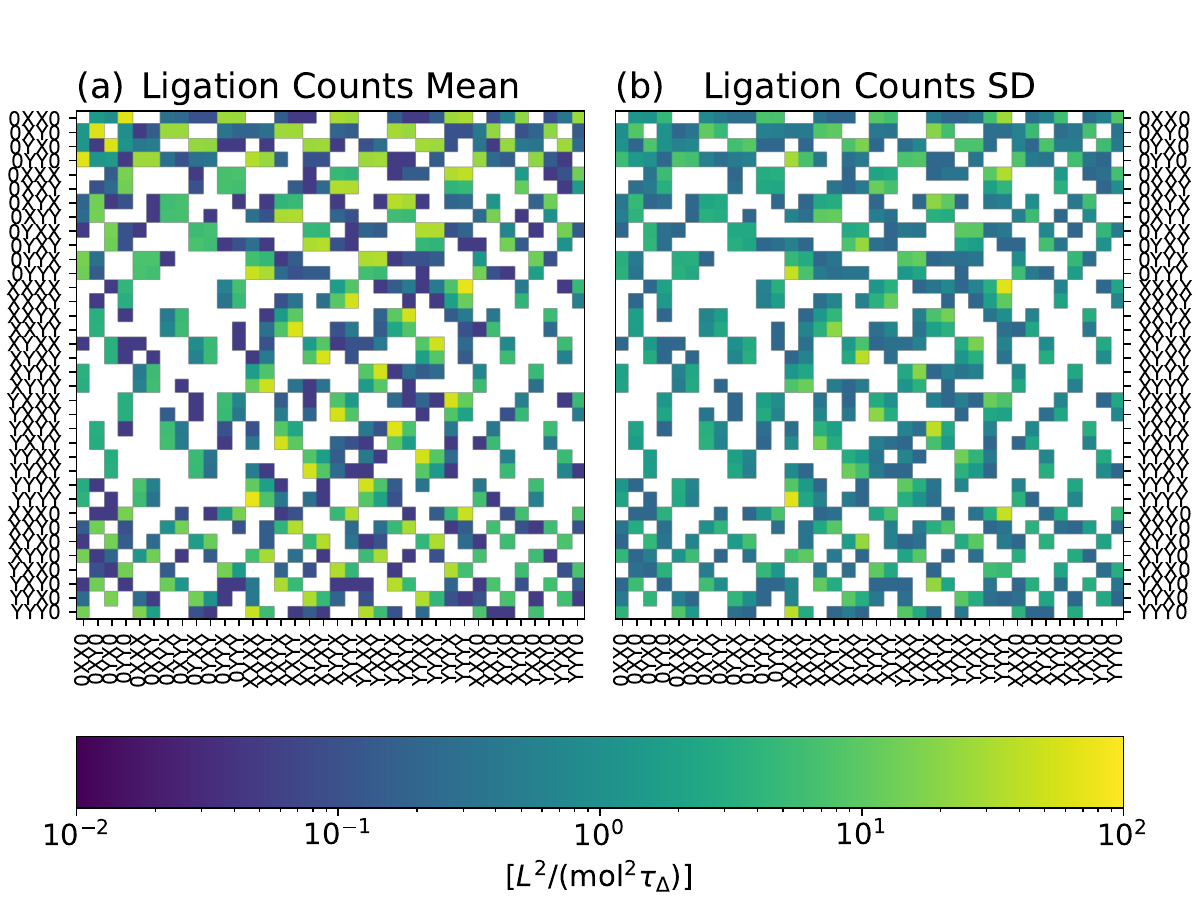}
    \caption{\label{fig:LigationCountsI}
    (a) Ligation counts mean and (b) standard deviation of parameter
    set~1 that serve as the data for the rate constants inference in scenario~1.
    Counts of zero are shown in white.
    }
\end{figure}

\begin{figure}
    \centering
    \includegraphics[width=\columnwidth]{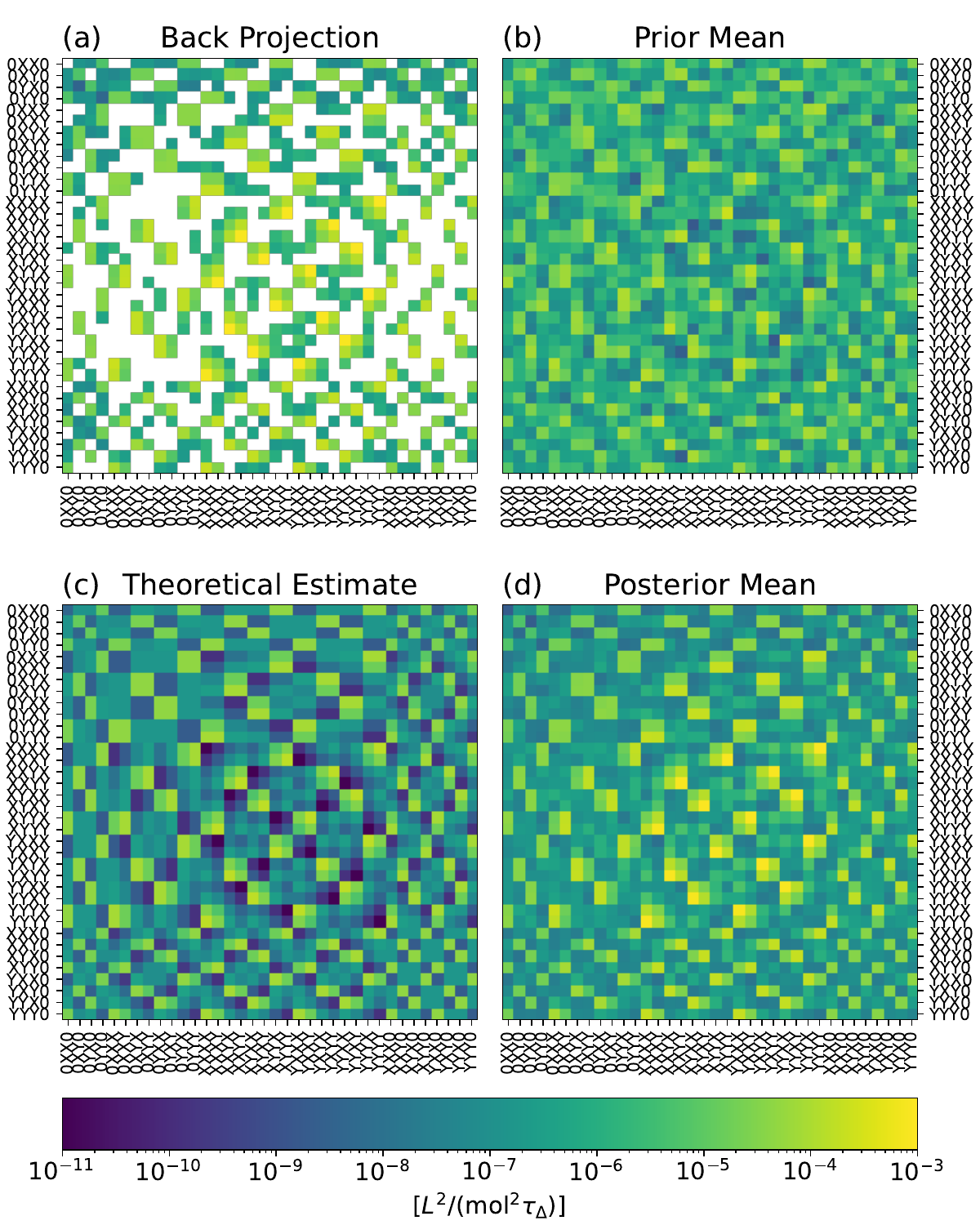}
    \caption{\label{fig:MeansI}
    (a) Back projected ligation counts,
    (b) prior mean rate constants,
    (c) theoretical mean rate constants, and
    (d) posterior mean rate constants
    for scenario~1.
    }
\end{figure}

First, we use ligation counts of twenty strand reactor simulation trajectories
for parameter set 1, see Fig.~\ref{fig:LigationCountsI}.
Solving Eqn.~\eqref{eq:rrc} for the motif extension rate constants by
dividing the expected motif extension counts through the reaction channel
exposures, we get the back projection, $\empc_\reprod/\left( \vol \mpe_\reprod \right)$, as purely descriptive estimate of the
motif extension rate constants, see~Fig.~\ref{fig:MeansI}a.
As prior rate constants, we use Eqn.~\eqref{eq:MPRC} for parameter set 0 without
hybridization bias and stalling (Fig.~\ref{fig:MeansI}b).
Parameter set 1 additionally considers ligation stalling in the presence of
mismatches according to Eqn.~\eqref{eq:Stalling}.
From those parameters, motif extension rate constants can be computed
theoretically according to Eqn.~\eqref{eq:MPRC}, see Fig.~\ref{fig:MeansI}c.
Note that those cannot be considerd as the true motif extension rate constants
as several assumptions and simplifications influenced our derivation:
First, the assumption of a steady hybridization-dehybridization state
that is computed in detail in the strand reactor simulation;
Second that reduction of motif extension rate constants to 4-mers that do not
consider longer hybridization sites.
This tends to an underestimation of the hybridization energy,
thus an overestimation of the dehybridization and in the end an underestimation
of the motif extension rate constants for later simulation times,
where longer strands dominate the population~\cite{HarthKitzerow2026Sequence}.
Still, it gives a good standard for judging the quality
of the inferred (posterior) motif extension rate constants,
using geoVI as described in Section~\ref{sec:GeoVI}.
In fact, the posterior motif extension rate constants mirror the structure of
the theoretical estimate at least qualitativly, see Fig.~\ref{fig:MeansI}d.
As expected, their mean is higher, since longer strands in later simulation times
increase the hybridization energy as opposed to the theoretical values that
only consider tetramers and shorter oligomers.

\begin{figure}
    \centering
    \includegraphics[width=\columnwidth]{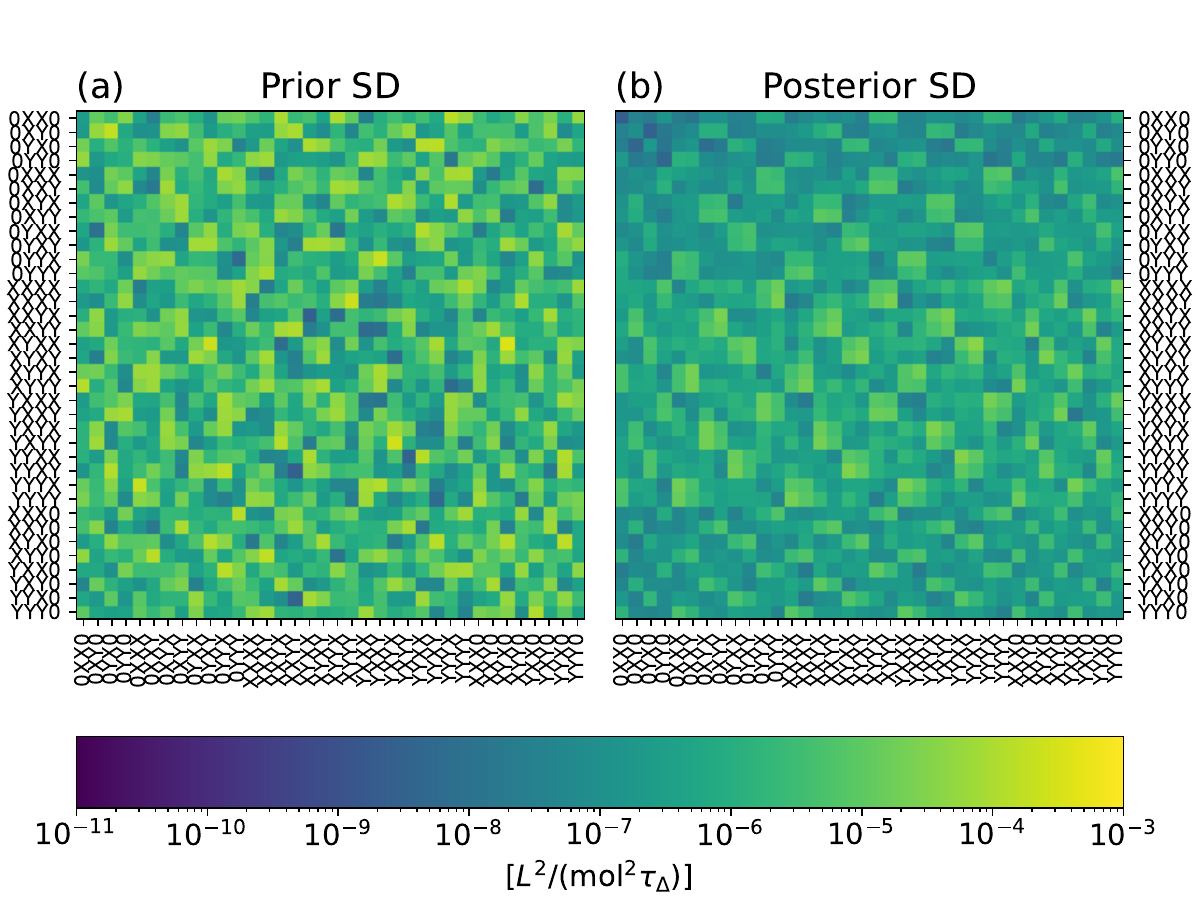}
    \caption{\label{fig:StdI}
    (a) Prior and (b) posterior rate constants standard deviations
    in scenario~1.
    }
\end{figure}

Additionally to the reaction rate constants mean,
we get covariances and uncertainties.
GeoVI is a Variational Inference method that provides posterior samples
after the minimization starting from a set of prior samples.
The prior standard deviations are set at the order of $10^{-4}
\frac{\mathrm{L}^2}{\mathrm{mol}^2\timeunit}$, see Figure~\ref{fig:StdI}a.
This allows flexible adjustment during the inference following the data.
Thus, after the minimization, the covariance is reduced by several orders of
magnitude (Fig.~\ref{fig:StdI}b).

\begin{figure}
    \centering
    \includegraphics[width=.48\columnwidth]{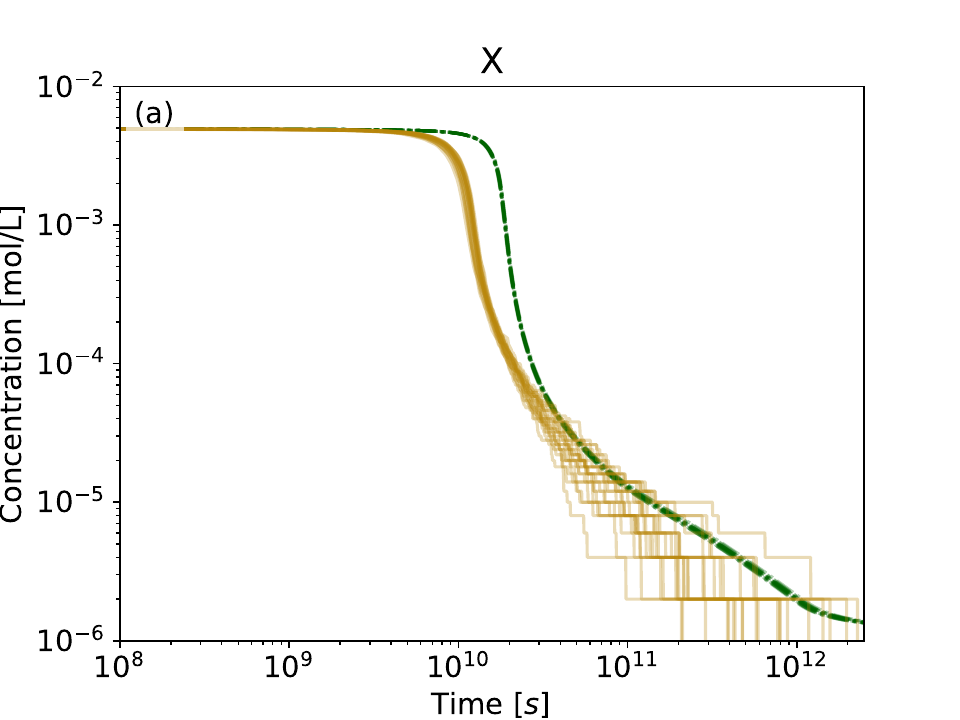}
    \includegraphics[width=.48\columnwidth]{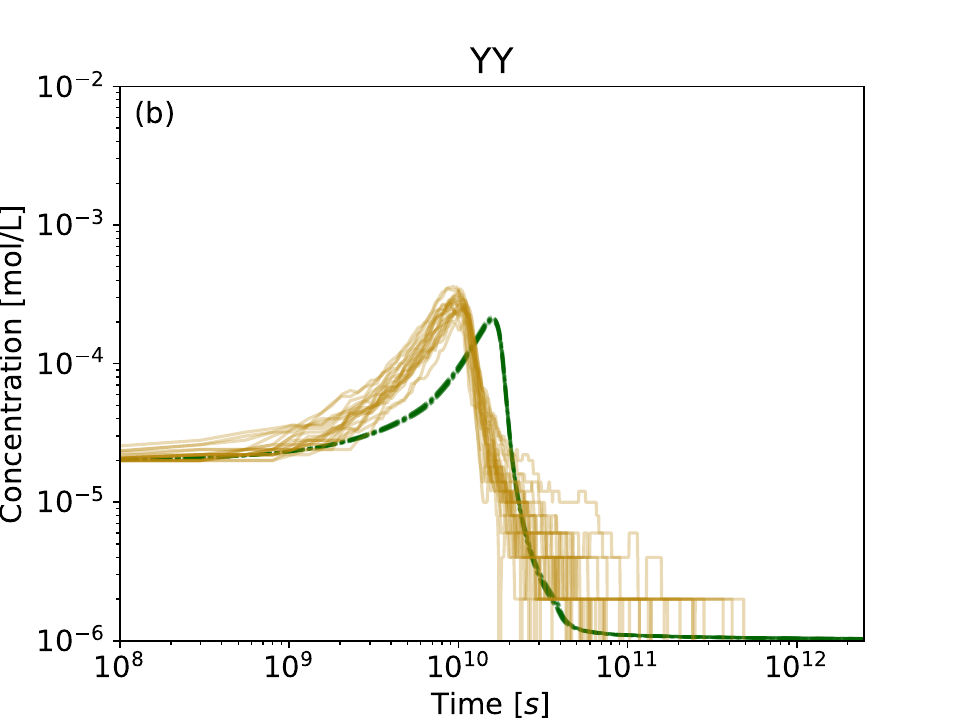}
    \includegraphics[width=.48\columnwidth]{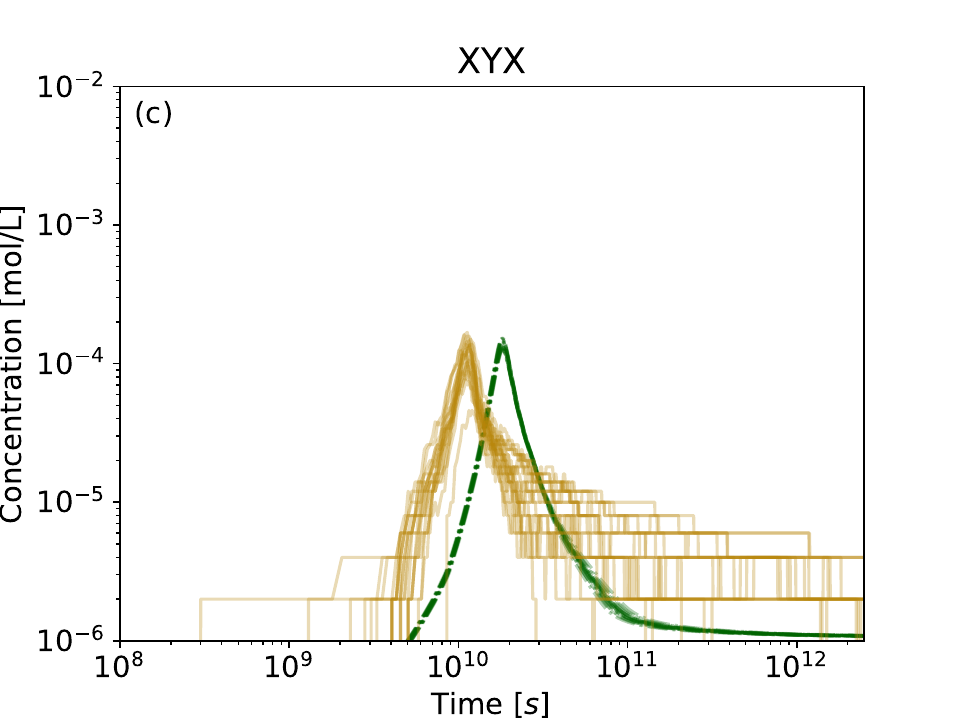}
    \includegraphics[width=.48\columnwidth]{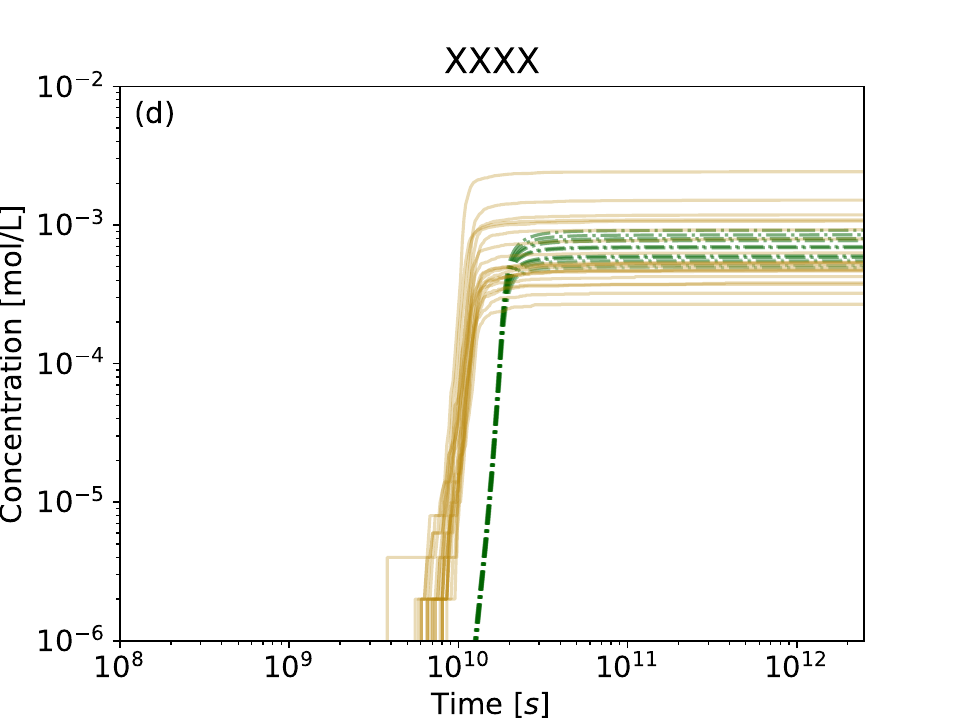}
    \caption{\label{fig:ConcentrationsI}
    Sequence motif concentrations of the original simulation (gold) and the
    integrated motif rate equations using inferred a posteriori rate constant
    samples (green) in scenario~1, for
    (a) monomers X,
    (b) dimers YY,
    (c) beginnings XYX, and
    (d) continuations XXXX.
    }
\end{figure}

Using the \texttt{MoRSAIK}-Package~\cite{HarthKitzerow2025MoRSAIK},
we can integrate the motif rate equations for each posterior sample and
directly compare the resulting motif concentration trajectories to the one simulated in the strand
reactor, see Fig.~\ref{fig:ConcentrationsI}.
Those trajectories reproduce the strand reactor trajectories qualitativly.
This confirms our reconstruction.
Effects like missing stochasticity, or finite pool effects and the reduction of strands
to shorter motifs are as expected~\cite{HarthKitzerow2026Sequence}.
Still, the motif extension rate constants tend to underestimate the kinetics
of the strand reactor simulation speedwise.
In the beginning and towards the end they agree perfectly.
The posterior trajectories even partially capture the variance in the
trajectories of the strand reactor simulation.
The reason behind this variance is completely different, though.
While in the strand reactor simulation the variance is an effect of
stochasticity in a finite system,
it is an uncertainty in the inferred parameters in case of the motif rate
equations.
This a posteriori uncertainty nevertheless is directly connected to the
variance in the ligation counts that stem from the same simulations like the
concentration trajectories.

\subsection{Hybridizaiton Bias}\label{sec:HybBias}

\begin{figure}
    \centering
    \includegraphics[width=\columnwidth]{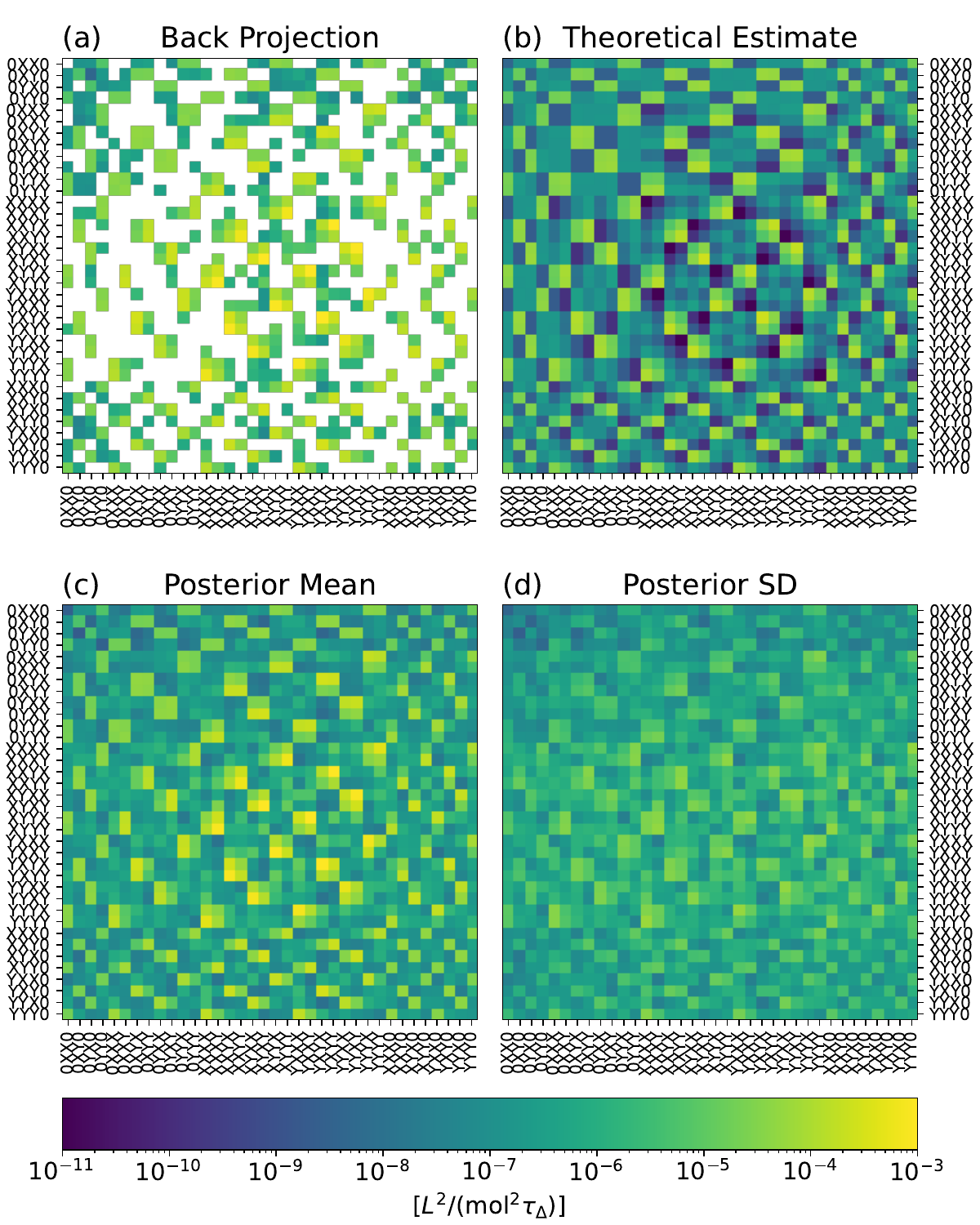}
    \caption{\label{fig:MeansII}
    (a) Back projected ligation counts,
    (b) theoretical mean rate constants,
    (c) posterior mean rate constants, and
    (d) posterior standard deviation
    for scenario~2.
    }
\end{figure}

Next, we infer motif extension rate constants from templated ligation counts
from strand reactor simulations with hybridization energy parameters
that discriminate between alternating and homogeneous sequences,
see~Section~\ref{sec:Evidence}.
Ligation stalling is set to the same value as before ($\stalparI=1,\stalparII=0.05$).
Using the same prior as before, the motif extension rate constants find similar
values to the theoretical estimations, see~Figure~\ref{fig:MeansII}.
Again the posterior mean rate constants tend to slightly higher values,
which is expected due to longer hybridization sites not being considered.

\begin{figure}
    \centering
    \includegraphics[width=.48\textwidth]{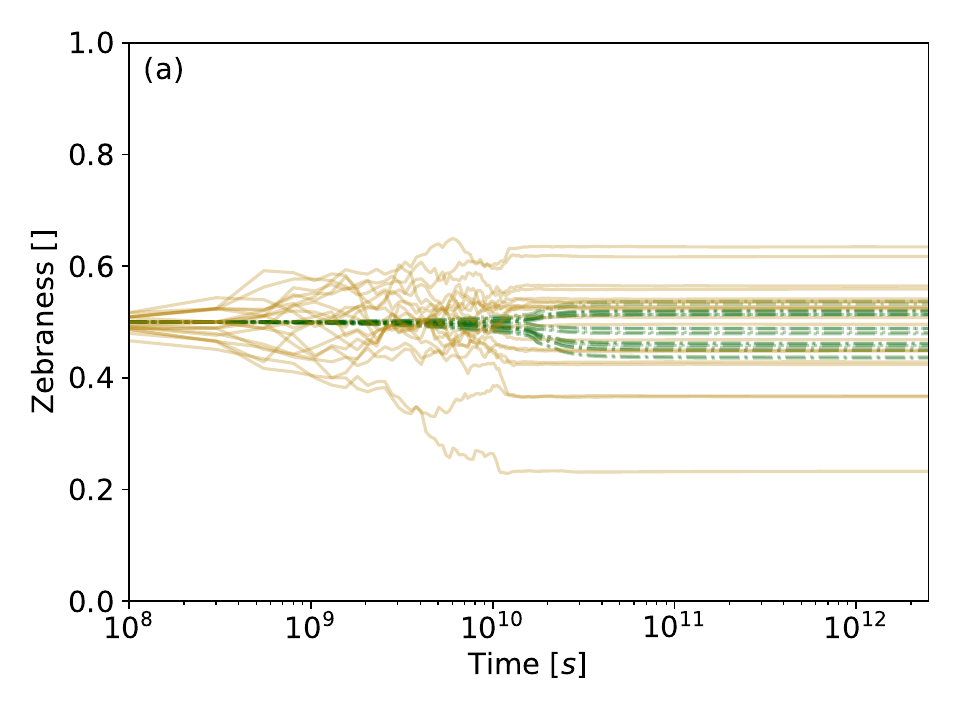}
    \includegraphics[width=.48\textwidth]{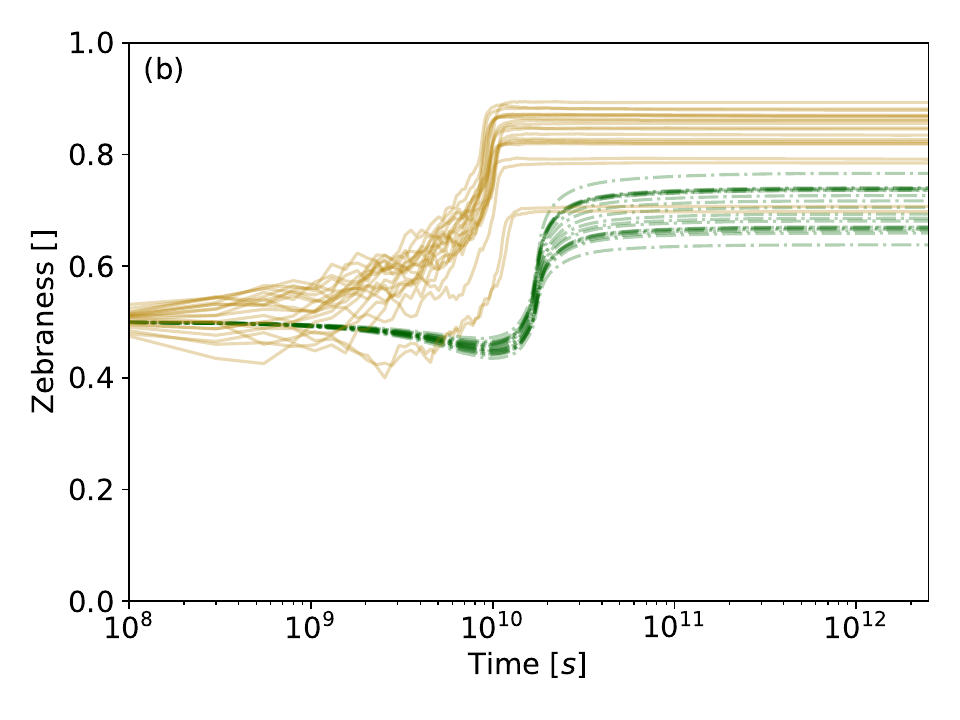}
    \caption{\label{fig:ZebranessII}
    Zebraness of the strand reactor simulation (gold) and the
    integrated motif rate equations using inferred a posteriori rate constant
    samples (green) (a) without and (b) with hybridization bias on alternating sequences.
    }
\end{figure}

Additionally, we can integrate motif rate equations using the inferred motif
extension rate constants.
As in previous work, we can compute the zebraness of the pool,
which is the fraction of alternating 2-mers
for a binary alphabet $\alphabet = \{\letterI,\letterII\}$~\cite{HarthKitzerow2026Sequence}.
It is 0.5 for a heterogeneous pool of as many homogeneous as alternating
sequences, 0 for a completely homogeneous sequence pool and
1 for only alternating sequences in the pool.
\begin{align}
    \slz = \frac{
        \sum_{\nuclvar_1,\nuclvar_4 \in \alphabet_0, \atop
        \motifvar_2, \motifvar_3 \in \alphabet}
        \c_{(\nuclvar_1,\motifvar_2,\motifvar_3,\nuclvar_4)}
        \cdot
        (1-\delta_{\motifvar_2, \motifvar_3})
    }{\sum_{\nuclvar_1,\nuclvar_4 \in \alphabet_0, \atop
        \motifvar_2, \motifvar_3 \in \alphabet}
        \c_{(\nuclvar_1,\motifvar_2,\motifvar_3,\nuclvar_4)}
    }\;,
\end{align}
where $\alphabet_0 := \alphabet \cup \left\{\0\right\}$, with $\0$ indicating
an empty spot for capturing beginnings, ends of strands.
In opposite to the sequence unbiased parameter set before, we now observe a
trend towards alternating motifs in the zebraness,
see~Figure~\ref{fig:ZebranessII}.
Both, the strand reactor trajectories and the integrated posterior motif rate
equation overlap, reassuring the reconstruction.
Still, the integrated motif rate equation tend to underestimate the zebraness
towards the end, which might be an effect of the unbiased prior on the motif
extension rate constants still present to some degree after the inference.

\section{Conclusion}

In this study, we presented and investigated a Bayesian inference framework
for nucleic acid reactors that perform hybridization, dehybridization and templated ligation.
For that we have projected the dynamics onto sequence motif space,
where they can be described by motif rate equations~\cite{HarthKitzerow2026Sequence}.
Using extensive strand reactor simulations~\cite{Rosenberger2021SelfAssembly,Goeppel2022Thermodynamic},
we generated templated ligation counts data from two different parameter sets
that consider ligation stalling and eventually a hybridization energy bias
for alternating sequences.
We have then used those parameters with different values to set the mean of our prior distribution
to infer motif extension rate constants from the generated data
using geometric Variational Inference~\cite{Frank2021GeoVI}.
Those agreed at least qualitatively with theoretical estimates from the
parameters of the underlying strand reactor simulations.
Integrating the motif rate equations with the inferred motif extension rate
constants confirmed that exact agreement between the
motif rate equations and the simulated strand reactor trajectories.

The Bayesian inference framework that we presented
can be used to calibrate motif rate equations to
extensive strand reactor simulations.
Those motif rate equations then allow to efficiently scan huge parameter spaces to get rough
estimates of the interesting reaction parameters before simulating them in detail with the strand
reactor simulation.
Furthermore, this is a first step towards Bayesian inference of reaction rate
constants from experimental data that capture the concentrations of
sequences at certain time points but not the reactions~\cite{Serrao2024Replication}.
In addition, recent developments on differentiable stochastic solvers~\cite{Burger2026Gradient} could be combined with the inference methods
presented here to build stochastic system inference methods
that are also capable of tracking 
finite size effects and stochasticity, which were shown to be crucial
for effects like spontaneous symmetry breaking in sequence
space~\cite{Goeppel2022Thermodynamic,HarthKitzerow2026Sequence}.

\section*{Acknowledgements}

This project was supported by the Deutsche Forschungsgemeinschaft (DFG, GermanResearch Foundation) under Germany’s Excellence Strategy – EXC-2094– 390783311.
Also, we thank
Bernhard Altaner,
Ludwig Burger,
Tobias G{\"o}ppel,
Julio Cesar Espinoza Campos,
Joachim Rosenberger,
Julian R{\"u}stig,
Philipp Frank,
Gordian Edenhofer,
Vincent Eberle,
Matteo Guardiani,
Viktoria Kainz,
Jakob Roth,
Paul Nemec,
Eike Eberhard,
%Julius Lehmann,
Julius Lehmann
and
Stephan Kremser
for fruitful discussions.

\section*{Author Contributions}

J.H.-K. performed the research.
J.H.-K., T.A.E. and U.G. designed the research.
J.H.-K. wrote the paper with input from all authors.

\medskip

\bibliographystyle{unsrt}
\bibliography{packages/bib}

\end{document}